\documentclass[conference,9pt]{IEEEtran}
\usepackage{algorithm}
\usepackage[noend]{algpseudocode}
\usepackage{caption}

\usepackage[T1]{fontenc}
\usepackage{amsmath,amssymb,mathtools}
\usepackage{booktabs,multirow,array,tabularx}
\usepackage{graphicx,xcolor}
\usepackage{tikz}
\usetikzlibrary{arrows.meta,positioning,fit}
\usepackage{cite}
\usepackage{balance}
\usepackage{url}
\usepackage{microtype}
\usepackage[hidelinks]{hyperref}
\hypersetup{pdfauthor={},pdftitle={SILK: Streaming Authentication with In-Place Checks for Quantized LLM Inference}}
\newcommand{\silk}{{SILK}}

\begin{document}

\newcommand{\cheng}[1]{{\color{blue} #1}}

\title{SILK: Streaming Authentication with In-Place Checks for Low-bit LLM Inference}
\author{
{\rm Ruichen Qi, Xinting Jiang, Junyi Luo, Ema Dimitrova, Quan Cheng, Mehdi Saligane}\\
Brown University
}
\maketitle


\begin{abstract}
A root-of-trust (RoT) supports model authentication at load time but does not guarantee integrity during inference. We present SILK, a quantization-aware in-place integrity framework protecting low-bit large language model (LLM) weights, activations, and KV caches against runtime tampering by embedding keyed checks directly into quantized data. Weight checks are embedded during second-order post-training quantization (GPTQ) to minimize distortion, while runtime tensors use redundant codes or fallback least significant bits (LSBs). An Ascon-based authentication chain absorbs each unit's cover bits into the next unit's authentication state, enabling single-pass verification. A streaming guard permits computation after bounded-strength confirmation and holds results until full authentication. Across evaluated symmetric INT4 models, mean perplexity increases remain below 0.4\% with 0.2\% weight metadata. Against Dual MAC guards, SILK cuts time overhead on Qwen2.5-0.5B weight reads at 4 B/cycle from 1.95\% to 0.41\% (Dual MAC-256) and energy-delay product (EDP) overhead at 4–8 B/cycle by 28–69\% with DDR (up to 37\% with HBM2), with 28\% more guard-logic area.

\end{abstract}

\begin{IEEEkeywords}
LLM accelerator, edge AI, in-place authentication, quantization, streaming integrity
\end{IEEEkeywords}
\vspace{-10pt}
\section{Introduction}
\label{sec:intro}


A root-of-trust (RoT) can authenticate a model at load time but does not guarantee integrity during inference. Low-bit large language model (LLM) accelerators repeatedly move model weights, activations, and KV caches through memory. When this memory and its interconnects are untrusted, data authenticated at load time can still be modified before use. Attacks on model weights~\cite{yao2020deephammer,rakin2019bit} and the need to protect intermediate tensors~\cite{ordonez2024enhancing} motivate integrity checks throughout inference. These checks must follow the accelerator's data accesses while limiting added traffic, buffering, and computation stalls.

Message authentication codes (MACs) can be used to provide runtime protection but introduce a granularity tradeoff. Fine-grained checks allow data to be verified as they arrive but require frequent tag transfers. Coarse-grained checks reduce metadata traffic and increase the amount of data covered by each verification. Under verify-before-use admission, an accelerator buffers the authentication block and waits for its check before starting computation~\cite{lee2023secureloop}. Overlapping verification with computation requires controlling the effects of tentative data until authentication completes~\cite{xuan2025seda,han2024tensortee}. The design must therefore specify when data may drive computation, dependent requests, and external writes.

\begin{figure}[t]
\centering
\includegraphics[width=\columnwidth]{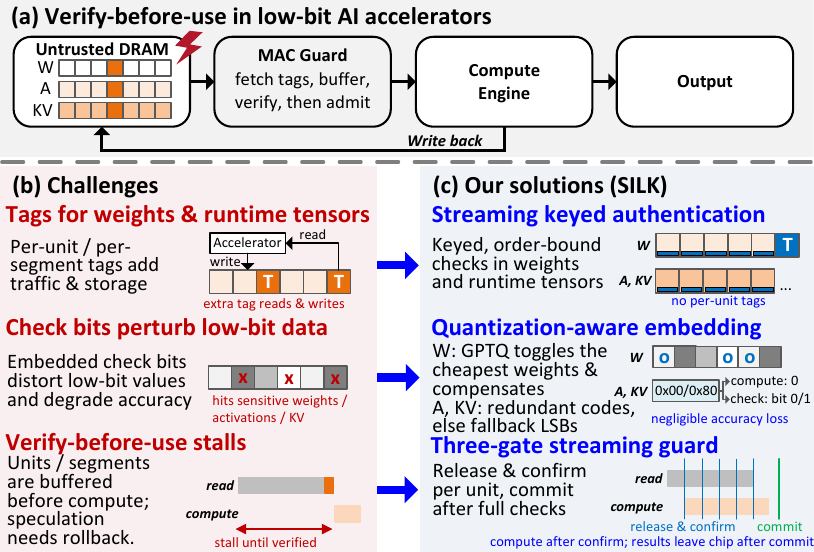}
\caption{Authentication in a low-bit accelerator. (a) A MAC guard reads
separate tags and verifies each block before admission. (b) Three challenges of verifying low-bit data before use and (c) SILK embeds checks in quantized tensors, confirms units as the stream arrives, and holds output writes until full authentication.}
\label{fig:intro}
\end{figure}


Quantization provides an opportunity to carry checks inside the protected data. Weight rounding offers a choice of which least-significant bits (LSBs) to modify, and redundant runtime codes can represent the same value with different check bits. Embedding a check, however, changes the data being authenticated. \silk{} coordinates embedding and verification to resolve this circular dependency. GPTQ~\cite{frantar2022gptq}-integrated constraints select low-cost weight LSB flips and compensate their error. A keyed chain generates each unit's check before absorbing that unit's modified cover bits with the next unit. This delayed absorption supports one-pass streaming verification.

\silk{} combines three key components (Fig.~\ref{fig:intro}).
\begin{itemize}\setlength{\itemsep}{0pt}\setlength{\parskip}{0pt}\item \textbf{Streaming keyed authentication.} An Ascon~\cite{nist2025ascon}-based chain uses delayed absorption to bind checks to data content, position, and order. Embedded unit checks support early confirmation, while weight-segment tags and runtime stream authentication govern commit.\item \textbf{Quantization-aware embedding.} GPTQ-integrated constraints embed weight checks with mean perplexity increases below 0.4\% across the evaluated symmetric INT4 models. Runtime checks use redundant codes and fallback LSBs, with fixed candidate positions for activations and calibrated channels for KV caches.\item \textbf{Three-gate streaming guard.} An advanced eXtensible interface (AXI) bridge admits computation and dependent reads after bounded-strength confirmation and holds writes until full authentication. On Qwen2.5-0.5B whole-model weight reads at 4~B/cycle, SILK adds 0.41\% execution time versus 1.95\% for Dual MAC-256; against Dual MAC guards, it reduces energy-delay product (EDP) overhead at 4--8~B/cycle by 28--69\% with DDR (up to 37\% with HBM2), with 28\% more guard-logic area.
\end{itemize}

\vspace{-10pt}
\section{Related Work and Threat Model}

\label{sec:model}

\subsection{Authenticated Data Admission in Secure Accelerators}
Authentication granularity affects both metadata traffic and computation stalls. SecureLoop jointly optimizes authentication block sizes and deep neural network (DNN) scheduling~\cite{lee2023secureloop}, while other designs explore multi-granular MACs, integrity trees~\cite{lee2025unified}, and multi-level verification~\cite{xuan2025seda}. Fine-grained MACs require separate tags. Under verify-before-use admission, coarse-grained MACs delay computation until entire segments are verified. Speculative execution can overlap verification with computation, but ensuring that unverified data have no irreversible effects complicates register-transfer level (RTL) verification~\cite{tan2025contract} and introduces side-channel leakage risks~\cite{cheng2024speclfb}.

\subsection{Integrity Protection for AI Accelerators}
GuardNN~\cite{hua2022guardnn} and other secure neural processing unit (NPU) designs~\cite{feng2024snpu,guardain2025,shen2022soter,xiong2022secndp} protect accelerator execution, while fine-grained bit-flip detectors~\cite{li2021radar,javaheripi2021hashtag} focus on detecting changes to model data. In-place error-correcting codes~\cite{guan2019place,nguyen2019st} target memory faults. Synergy~\cite{saileshwar2018synergy} co-locates MACs with data in ECC-DIMMs and uses separately stored parity for error correction.

Other techniques embed checks directly into data. E$^3$-CODE~\cite{kamineni2026e3} optimizes embedded parity placement across weight representations for resilience to burst errors. SANAI~\cite{rahaman2026sanai} embeds keyed error-correcting code (ECC) parity in model weights and uses keyed interleaving to disperse concentrated bit-flip attacks. Fragile and reversible watermarks~\cite{guan2020reversible,botta2021neunac} embed marks for model-integrity checking. Syndrome-trellis codes~\cite{filler2011stc} support low-distortion embedding with public constraints, while progressive MACs~\cite{armknecht2020promacs} strengthen authentication as subsequent tags arrive. SILK combines quantization-aware embedding with admission control for weights and runtime tensors, binding checks to authenticated contexts and their position in a stream (Table~\ref{tab:related}).

\subsection{Threat Model}

\silk{} protects data stored or transferred outside the trusted chip boundary. On-chip compute engines, buffers, the integrity guard, and hardware RoT are trusted, while external memory, controllers, interconnects, and direct memory access (DMA) paths are untrusted. Attackers may read, modify, reorder, replay, or truncate data. A trusted toolchain embeds weight checks offline, while the guard embeds runtime checks on writes and verifies them on reads, with trusted descriptors binding data to regions, formats, and versions. Like SecureLoop~\cite{lee2023secureloop}, \silk{} prevents unverified data from causing irreversible effects: computation and dependent reads require bounded-strength confirmation, while outputs, memory writes, and updates to stored stream records require full authentication. Pending results are discarded on failure without rollback, and a trusted counter limits attempts per key slot (Section~\ref{sec:security}). \silk{} protects integrity, not confidentiality; malicious authorized models, physical attacks, side channels, and denial of service are out of scope. ECC-corrected errors do not count toward the failure limit.

\begin{figure}[t]
\centering
\includegraphics[width=\columnwidth]{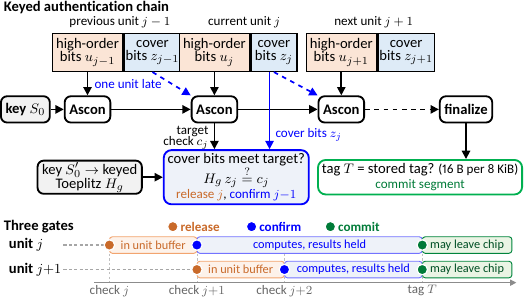}

\caption{Top: Keyed authentication chain for a weight segment. Each unit contains high-order bits $u_j$ and cover bits $z_j$, absorbed one unit late; its check passes when $H_gz_j=c_j$ (Eq.~\eqref{eq:coset}). Bottom: At $N=1$, passing unit $j$ releases itself and confirms unit $j-1$. Confirmed units may enter computation, but results remain on chip until the final segment tag $T$ verifies.}

\label{fig:chain}
\vspace{-8pt}

\end{figure}

\begin{figure}[t]
\centering
\includegraphics[width=\columnwidth]{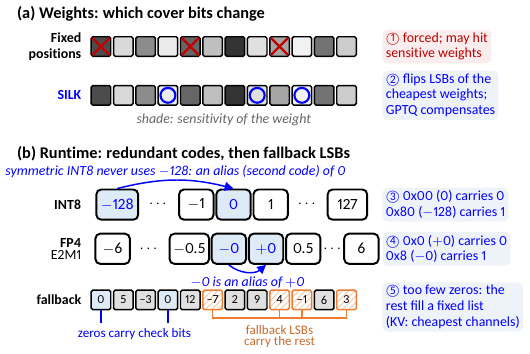}
\caption{Carriers: (a) a dense keyed $H_g$ lets the embedder flip the cheapest cover bits; (b) runtime checks use redundant codes (INT8 restricted to $[-127,127]$; FP4 for NVFP4 and MXFP4), then fallback LSBs (last $\ell$ elements or, for KV, calibrated channels).}
\label{fig:carriers}
\vspace{-8pt}
\end{figure}

\begin{figure*}[t]
\centering
\includegraphics[width=\textwidth]{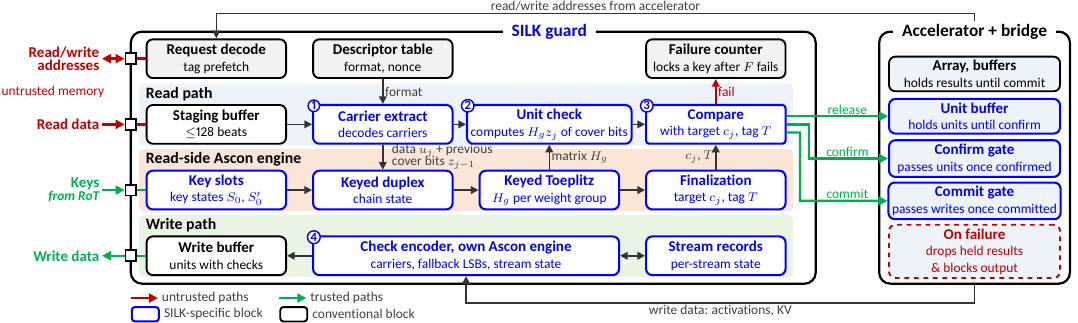}
\caption{SILK guard architecture. \textcircled{\scriptsize 1} Carrier extraction recovers runtime check bits and restores primary (non-alias) codes; \textcircled{\scriptsize 2} each unit's embedded check (cover or carrier bits) is compared with the chain's target check, which absorbs cover bits one unit late; \textcircled{\scriptsize 3} release, confirm, and commit events drive the AXI bridge (Table~\ref{tab:gates}); \textcircled{\scriptsize 4} a separate write-path engine embeds checks into activations and KV.}

\label{fig:guard}
\vspace{-8pt}
\end{figure*}

\section{SILK Architecture}
\label{sec:architecture}
In this section, we first introduce the keyed authentication chain and its delayed-absorption mechanism, followed by check embedding methods for quantized weights and runtime tensors. We then describe the streaming guard hardware and analyze its security guarantees.

\subsection{Keyed Authentication Chain with Delayed Absorption}
\label{sec:construction}
\silk{} divides weights, activations, and KV caches into 256-byte \emph{units} (512~B for INT8 weights, so every weight unit has 512 cover bits; Fig.~\ref{fig:chain}). For INT4 weights, the 512 code least significant bits (LSBs) form the \emph{cover bits} $z_j$ of unit $j$ (indexed from zero within each chain), while its other bits form $u_j$. Weight chains span 32 units (8 KiB for INT4, 16 KiB for INT8); runtime chains follow individual activation, key, or value streams as they grow.

Each chain uses a keyed Ascon duplex with a 320-bit state and a 128-bit key. A trusted header binds the region, segment or stream, nonce, and data format, as well as the generation and key epoch for runtime streams. The 12-round Ascon permutation processes 32-byte blocks in order and produces a target check $c_j$ of length $\ell$ bits for each unit. At most 128 state bits are exposed, leaving 192 hidden. The chain instantiates the generic full-state keyed duplex~\cite{daemen2017duplex,mennink2023duplex} with the 12-round Ascon permutation rather than a standardized Ascon mode, which outputs one tag per message; its security thus rests on that generic analysis (Section~\ref{sec:security}).

Embedding a check would normally change the data used to generate it. \silk{} breaks this circular dependency through \emph{delayed absorption}: to generate $c_j$, the chain absorbs $u_j$ and the previous cover bits $z_{j-1}$, but not $z_j$. The new $z_j$ enters with the next unit (Fig.~\ref{fig:chain}). Finalization absorbs the last cover bits and the unit count, ensuring that every stored bit affects the final authentication state without a second pass.

For each group $g$ of weight units, a secret, dense Toeplitz matrix $H_g$, derived from a separate key domain, defines the binary check constraint:
\begin{equation}
\label{eq:coset}
H_g z_j=c_j,\qquad H_g\in\{0,1\}^{\ell\times512}.
\end{equation} Its many valid solutions allow low-distortion embedding, while keeping the matrix secret and dense makes passing alterations harder than under public or sparse constraints. Weight units use $\ell=36$; runtime units use 32-bit checks carried by redundant codes or fallback LSBs. Each weight segment ends with a 128-bit tag, whereas runtime streams keep a trusted on-chip stream state.

The checks open three gates (Table~\ref{tab:gates}). A unit is \emph{released} after its own check passes: it advances the chain and waits in the unit buffer. Weight units are \emph{confirmed} for computation only after the next $N$ units also pass, where $N$ is the confirmation depth; the trailing units instead require the final segment tag. A change to unit $j$ alters the chain state by unit $j+1$, making the following $N$ checks unpredictable and adding roughly $\ell$ bits of confirmation strength per check before accounting for repeated attempts. \emph{Commit} allows results to leave the chip only after full authentication of the data they depend on: the segment tag for weights; for runtime units, which confirm at their own check, five further checks or a trusted on-chip stream state match, covering the stream up to that point but not later appends.

\subsection{Quantization-Aware Check Embedding}
\label{sec:weights}
\silk{} embeds weight checks during GPTQ rather than after quantization, so later columns can compensate for embedding error. A unit is a 512-column slice of one weight row (Fig.~\ref{fig:carriers}a). Let $q$ be its rounded LSB vector and $b$ mark the LSBs selected for flipping. The stored cover bits become $z_j=q\oplus b$, which must satisfy $H_gb=c_j\oplus H_gq$. These flips leave the high-order code bits unchanged.

The embedder seeks a valid set of flips with low quantization loss. Let $V_{ii}$ be the signed change in weight $i$ caused by flipping its code LSB. With diagonal matrix $V$, the embedding perturbation is $\Delta w=Vb$. With $\Sigma$ the layer Hessian, $P$ the current unit's columns, $R$ the later columns, and $\gamma$ the gradient from previously committed quantization errors, the loss after compensating later columns is
\begin{equation}
\label{eq:schur}
\mathcal{L}(\Delta w)=\Delta w^{\top}M\Delta w+2\gamma^{\top}\Delta w,\quad M=\Sigma_{PP}-\Sigma_{PR}\Sigma_{RR}^{-1}\Sigma_{RP}.
\end{equation}
Writing $W_P$ and $W_R$ for the weights in columns $P$ and $R$, Algorithm~\ref{alg:embed} summarizes the process: find low-cost flips that satisfy the check, refine them without changing the check, and compensate their error in later columns. Embedding is performed once per key and takes about twice the time of GPTQ alone. For other weight formats, it uses INT8 code LSBs or NVFP4/MXFP4 mantissa LSBs; scales and zero points are protected in a separate FP16 region.


\begin{algorithm}[t]
\caption{Quantization-Aware Check Embedding}
\label{alg:embed}
\small
\begin{algorithmic}[1]
\Require weights $W$, Hessian $\Sigma$, key $K$
\State $m\gets96$ \Comment{candidate positions}
\For{each weight segment $s$}
  \State initialize chain from $K$ and segment header
  \For{each unit $j$ in $s$ (columns $P$, later $R$)}
    \Statex \textbf{Stage 1: Target check generation}
    \State $(q,u_j)\gets$ GPTQ-round $W_P$
    \State absorb $u_j$ and, if $j>0$, $z_{j-1}$
    \State extract target check $c_j$

    \Statex \textbf{Stage 2: Low-cost bit selection}
    \State $d_i\gets V_{ii}^2M_{ii}+2V_{ii}\gamma_i$
           for each cover bit $i$
    \State $C\gets$ $m$ cheapest positions,
           sorted by $d_i$
    \State $H_{g,C}\gets$ columns of $H_g$ indexed by $C$
    \State $b_C\gets$ pivot solution of
           $H_{g,C}b_C=c_j\oplus H_gq$,
    \Statex \hspace{\algorithmicindent}
           eliminating columns in $C$ order
    \State $b\gets$ zero-padded $b_C$
    \State $\mathcal{B}\gets$ one zero-padded
           $\ker H_{g,C}$ vector per non-pivot column
    \For{$t=1,\ldots,40$}
      \State $k\gets\arg\min_{k'\in\mathcal{B}}
             \mathcal{L}(V(b\oplus k'))$
      \If{$\mathcal{L}(V(b\oplus k))\ge\mathcal{L}(Vb)$}
        \State \textbf{break}
      \EndIf
      \State $b\gets b\oplus k$
    \EndFor

    \Statex \textbf{Stage 3: GPTQ error compensation}
    \State $z_j\gets q\oplus b$
    \State $W_R\gets W_R-
           (Vb)^\top\Sigma_{PR}\Sigma_{RR}^{-1}$
  \EndFor
  \State absorb final $z_j$ and unit count
  \State output embedded weights and tag \(T_s\)
\EndFor
\end{algorithmic}
\end{algorithm}

\subsection{Runtime Redundant Codes}
\label{sec:runtime}

Unlike offline weights, runtime activations and KV caches use redundant codes to embed checks (Fig.~\ref{fig:carriers}b). For INT8 quantizers restricted to $[-127,127]$, \silk{} reserves code $-128$ as a zero alias; FP4 provides $+0$ and $-0$. These codes carry check bits without changing values; the reader extracts the check bits and then restores the primary codes (by our convention, those carrying 0) for computation.

Each region fixes an ordered list of $\ell$ fallback positions before embedding. Each unit uses its first $\ell$ zeros outside this list as carriers; with $n_c<\ell$ carriers, the remaining $\ell-n_c$ bits overwrite the LSBs of the first $\ell-n_c$ listed positions, changing each value by at most one LSB. For KV, the list holds the $\ell$ channels whose one-LSB change costs least, ranked cheapest first by its mean squared error weighted by query energy (keys) or output-projection column norm (values), calibrated offline on WikiText-2. Activation units start at different channels, so their list is simply the last $\ell$ elements. Formats without aliases, such as asymmetric INT8 and KIVI~\cite{liu2024kivi}, use fallback LSBs alone.

Shared carrier and fallback rules eliminate position metadata and online search, and the reader buffers at most $\ell$ LSBs. The chain absorbs carrier values and all other data bits, while the alias pattern and fallback LSBs hold the check compared with $c_j$, so any other alias pattern fails. Fallback errors are included in accuracy evaluation.

KV streams grow token by token. On-chip records retain each stream's generation, unit count, and chain state for appends and reads verified from the beginning; new generations prevent restart replay. When the table fills, cold records are protected off chip by Ascon authenticated encryption with associated data (AEAD) bound to key epoch, stream, generation, and count. A trusted directory rejects stale, duplicate, or foreign restores. Each record is 348 bits; the evaluated 1,024-record table (44~KiB) is counted in Table~\ref{tab:guards}, and no record was evicted in our timing runs.

\subsection{Streaming Guard Hardware}
\label{sec:guard}

Fig.~\ref{fig:guard} shows the guard between untrusted memory and the accelerator. The request decoder uses trusted descriptors to select the format, nonce, and key slot and prefetch weight-segment tags. RoT-provided keys remain on chip.

On reads, \textcircled{\scriptsize 1}~\emph{Carrier extraction} recovers runtime checks from aliases and fallback LSBs and restores primary codes; weights pass unchanged. \textcircled{\scriptsize 2}~\emph{Unit checking} computes $H_gz_j$ for weights or extracts runtime check bits, while the Ascon engine derives $c_j$ through delayed absorption. \textcircled{\scriptsize 3}~\emph{Comparison} matches these checks and verifies the final segment tag or trusted on-chip stream state when required. Failures abort the chain and update the key-slot failure counter.

The resulting \emph{release}, \emph{confirm}, and \emph{commit} events control the AXI bridge (Table~\ref{tab:gates}). Release buffers passing units; confirm permits computation and dependent reads. For an output depending on weight unit $j$ in segment $s$ and a KV unit in stream $r$, at $N=1$ the weight confirms after unit $j+1$ passes (or its final tag matches), while the KV unit confirms after its own check. The output remains on chip until both units commit; otherwise it is discarded without rollback. Writes wait per layer in the output buffer, which stalls the array when full; all reported times include these waits.

On writes, a separate \textcircled{\scriptsize 4}~\emph{Ascon engine} embeds checks into activations and KV data using the redundant codes and fallback LSBs of Section~\ref{sec:runtime}. A two-unit buffer holds one unit while the next is absorbed; per-stream records retain generation, unit count, and chain state. The encoded units are then written to untrusted memory.

\begin{table}[t]
\centering
\caption{Admission gates and nominal online false-acceptance bounds (Section~\ref{sec:security}).}
\label{tab:gates}
\scriptsize\setlength{\tabcolsep}{2.6pt}
\begin{tabular}{@{}lllcccc@{}}\toprule
 & & & \multicolumn{3}{c}{weights ($\ell{=}36$)} & runtime tensors\\
\cmidrule(lr){4-6}\cmidrule(l){7-7}
Gate & Opens when & Data may & $N{=}1$ & $N{=}2$ & $N{=}4$ & ($\ell{=}32$)\\
\midrule
Release & own check passes & wait in buffer
& \multicolumn{3}{c}{$2^{-36}$ \ ($2^{-27\,*}$)} & $2^{-32}$\\
Confirm & required checks pass & enter compute
& $2^{-30}$ & $2^{-61}$ & $2^{-122}$ & $2^{-26}$\\
Commit & full auth. passes & leave chip
& \multicolumn{3}{c}{$2^{-122}$} & $2^{-136}$\\
\bottomrule
\end{tabular}
\par\vspace{2pt}
\noindent\parbox{\columnwidth}{\raggedright\scriptsize

\emph{Notes:} Release is per attempt; confirmation is per key slot
($F=64$). Weight commit allows up to \(F=64\) forgery attempts (Eq. (4)). Runtime $2^{-136}$ bounds five further
32-bit checks ($N=5$, $F=64$, Eq.~\eqref{eq:confirm}), not
trusted on-chip stream state matching. $^*$ denotes the strongest
measured adaptive release attack. At $N{=}4$, non-trailing units reach $2^{-124}$; trailing units confirm by the tag.

}
\end{table}

\subsection{Security Analysis}
\label{sec:security}
We consider an attacker who controls off-chip data, observes embedded checks and tags, and submits modified data adaptively. The goal is to commit a result using data not authorized for its region, position, and version. Our analysis models Ascon as an ideal permutation~\cite{mennink2023duplex} and assumes trusted embedding, independent 128-bit device keys, fresh nonces, and at most one embedding per position and context. A trusted counter locks each key slot after $F=64$ failures. Online bounds are per key slot; success across re-keying or multiple devices requires aggregating their online attack budgets. Each chain is modeled as one session of a full-state keyed duplex~\cite{daemen2017duplex} with a 192-bit hidden state; keyed initialization, XOR-only absorption, and delayed absorption define its query interface. Let $\mu$ be the number of device keys and $Q_p$ the number of offline permutation queries. We analyze confirmation and commit in the ideal-permutation model; Theorem~1 of~\cite{mennink2023duplex} (no overwrites, one repeated subpath per session) bounds the duplex correction $\varepsilon_{\mathrm{KD}}$ for $L$ sessions (chains and $H_g$ derivations); its leading term is $LQ_p/2^{192}$, e.g., $2^{-83}$ for $L{=}2^{45}$ and $Q_p{=}2^{64}$, and its other terms total below $2^{-96}$ for $\mu\le2^{16}$.

\emph{Release.} A blind change must satisfy an unknown keyed check, but low-distortion embedding may leak information to an adaptive attacker. Release therefore never admits data to computation on its own.

\emph{Confirmation.} A change to unit $j$ puts the chain on a previously unembedded path by unit $j+1$ at the latest. Under the ideal-permutation model, its unqueried 192-bit hidden state makes later checks unpredictable even if $H_g$ is known. A change must pass $N$ fresh checks within $F$ failures, each no likelier than guessing an $\ell$-bit target; a union bound gives
\begin{equation}
\label{eq:confirm}
\Pr[\text{false confirm}]\le \tbinom{F+N}{N}\,(2^{\ell}-F)^{-N}+\frac{\mu Q_p}{2^{128}}+\varepsilon_{\mathrm{KD}}.
\end{equation}Table~\ref{tab:gates} reports the online (first) terms.

\emph{Commit.} Finalization binds every data bit and the authentication context. Let $q_v$ be the number of submitted chains. Under the same duplex model~\cite{mennink2023duplex},
\begin{equation}
\label{eq:commit}
\Pr[\text{false commit}]\le\frac{q_v}{2^{128}-q_v}+\frac{\mu Q_p}{2^{128}}+\varepsilon_{\mathrm{KD}}.
\end{equation}For weights, Table~\ref{tab:gates} reports a nominal \(2^{-122}\) bound over \(F=64\) attempts; a single-tag forgery has a nominal bound of \(2^{-128}\), while $2^{-136}$ bounds five further 32-bit runtime checks under $F=64$ via Eq.~\eqref{eq:confirm}. Runtime commit may alternatively require a match with the trusted on-chip stream state.

\vspace{-4pt}
\section{Evaluation}
\label{sec:result}

\subsection{Experimental Setup}
We evaluate GPTQ-quantized Qwen2.5 (0.5B--14B) and TinyLlama (TL) on WikiText-2 perplexity increase over unchecked quantization, six-task zero-shot accuracy, and 5-shot GSM8K. The guard RTL matches a Python reference bit-exactly and is synthesized in ASAP7~\cite{clark2016asap7}, with an AXI bridge to a 32$\times$32 INT8 systolic array~\cite{xuan2025seda}. RTL tests at 4--16~B/cycle cover Gemma~3 270M prefill and attention, full INT4 Qwen2.5-0.5B, and 1K/4K-token KV attention. Guard-level results report the worst of ten tiles at 4~B/cycle, with 16-B tags per 8-KiB INT4 or 16-KiB INT8 segment. Decode tests use INT4, one token per step, and a 16-KiB output buffer. Energy estimates combine gate-level power, 7-nm SRAM, and DDR/HBM2 traffic at 20.3/3.97~pJ/bit~\cite{jouppi2021ten,oconnor2017fine}.

MAC baselines match failure budgets, versioning, and stream support. Dual MAC-256/512 uses 32-bit unit checks plus a 16-B tag per 32 units; MAC-256/512 uses 64-bit unit checks and no segment tag, matching SILK's confirmation strength. The 512-B granularity follows a configuration evaluated in MGX~\cite{hua2022mgx}. Seg.\ MAC uses one tag per segment (8 KiB for INT4 and 16 KiB for INT8), verifying before use as in SecureLoop~\cite{lee2023secureloop} or computing before verification as in SeDA, TensorTEE, and Securator~\cite{shrivastava2023securator,han2024tensortee,xuan2025seda}.

\begin{figure}[t]
\captionsetup{justification=justified,singlelinecheck=false}
\centering
\includegraphics[width=\columnwidth]{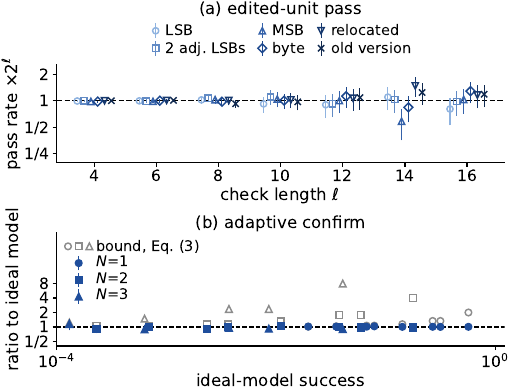}
\caption{Security of embedded checks: (a) edited-unit pass rates relative to ideal $2^{-\ell}$; (b) adaptive confirmation at reduced widths ($\ell=4$--8) versus Eq.~\eqref{eq:confirm} (hollow markers) and the ideal model. Deployment-width bounds ($\ell=36$ for weights, $\ell=32$ for runtime) are analytical.}
\label{fig:eval}
\end{figure}

\subsection{Security Evaluation}
\label{sec:res-sec}
We first test whether individual checks and the full chain provide the protection expected by the security analysis. Across six change types, unit-check pass rates track the ideal $\ell$-bit behavior (Fig.~\ref{fig:eval}a), using $10^5$--$2.6\times10^6$ keys per point and 95\% confidence intervals. A solver-aware attacker can gain up to 9 bits against release (Table~\ref{tab:gates}), but adaptive confirmation remains consistent with the analytical bound (Fig.~\ref{fig:eval}b). The latter uses reduced-width chains ($\ell=4$--8, $N=1$--3) to make successful attacks measurable. This explains why release alone cannot admit data to computation. The RTL tests then check whether the guard enforces the stronger gates: altered carriers or tags, replay, relocation, and incorrect keys or contexts are rejected or prevented from committing. Honest RTL runs match the Python reference bit-exactly. In the tested tampering cases, modified weights, activations, and KV units produce no output writes ($N=1$--4).

\begin{figure}[t]
\captionsetup{justification=justified,singlelinecheck=false}
\centering
\includegraphics[width=\columnwidth]{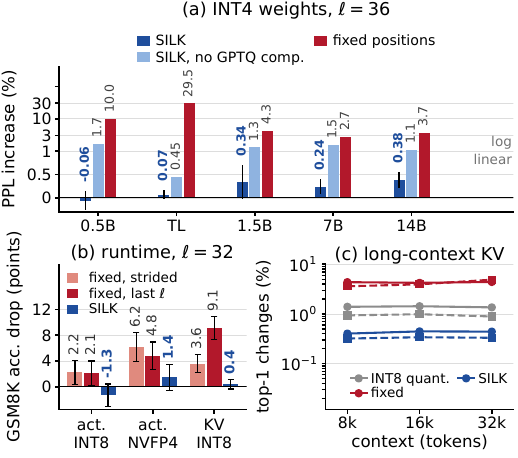}
\caption{Accuracy cost of embedded checks: (a) INT4 weight perplexity increase; (b) GSM8K accuracy drop (95\% paired bootstrap confidence intervals); (c) passkey prompts: greedy next-token changes relative to unchecked INT8 KV (INT8 quant.: unchecked INT8 vs.\ FP16 KV; solid: 1.5B; dashed: 7B).}

\label{fig:acc}
\end{figure}
\subsection{Accuracy of Weight Check Embedding}
\label{sec:res-acc}
Next, we assess the accuracy cost of weight checks. Fixed embedding uses predetermined bits, as in in-place ECC~\cite{guan2019place}. Figure~\ref{fig:acc}a reports means over eight SILK keys and one to six keys for the other methods; INT8 perplexity increases by at most 0.02\%. 
Integrating embedding into GPTQ lets low-cost LSB flips be compensated by later weights. On TinyLlama, the mean perplexity increase drops from 29.5\% with fixed positions to 0.45\% with SILK's bit selection alone, and to 0.07\% with GPTQ compensation (Fig.~\ref{fig:acc}a). Across models and keys, SILK keeps INT4 perplexity increase below 1\%, while fixed positions incur several to tens of percent.
Six-task zero-shot accuracy stays within 1.4 points, and other tested low-bit formats remain at ${\leq}0.7\%$ perplexity increase. GPTQ thus preserves accuracy, while the keyed chain provides integrity (Section~\ref{sec:res-sec}).

\subsection{Accuracy of Runtime Tensors}
\label{sec:res-rt}
\begin{figure}[t]
\captionsetup{justification=justified,singlelinecheck=false}
\centering
\includegraphics[width=\columnwidth]{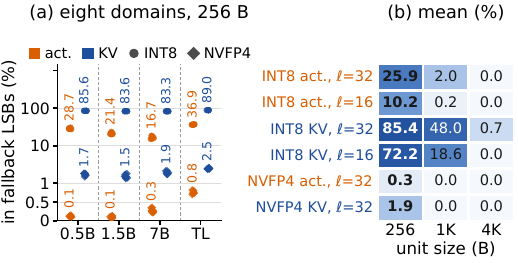}
\caption{Fraction of runtime check bits in fallback LSBs: (a) across eight domains ($\ell=32$; labels show means); (b) mean by unit size.}
\label{fig:fb}
\end{figure}

\begin{figure}[t]
\captionsetup{justification=justified,singlelinecheck=false}
\centering
\includegraphics[width=\columnwidth]{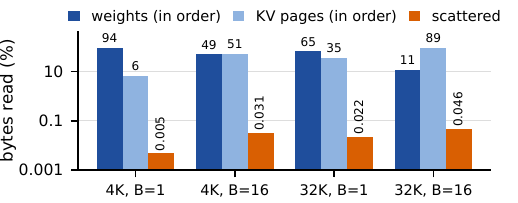}
\caption{LLM-decoding memory reads: in-order weights/KV (five-model mean) and scattered accesses (worst model); $B$: batch size.}
\label{fig:stream}
\end{figure}

\begin{table}[t]
\captionsetup{justification=justified,singlelinecheck=false}
\centering\caption{Guard-level metadata, time, energy-delay product (EDP), and hardware cost at comparable admission strength.}
\label{tab:guards}
\scriptsize\setlength{\tabcolsep}{1.45pt}
\begin{tabular}{@{}l|l|rr|r|rr|r|rr|r@{}}\toprule
 & & \multicolumn{5}{c|}{traffic and speed (\%)} & \multicolumn{4}{c}{cost}\\\cmidrule(lr){3-7}\cmidrule(l){8-11}
Guard & admit & \multicolumn{2}{c|}{meta} & time & \multicolumn{2}{c|}{EDP} & buf. & area & +SRAM & power\\
 & & INT4 & INT8 & & DDR & HBM2 & (KiB) & \multicolumn{2}{c|}{(k\textmu m$^2$)} & (mW)\\\midrule
\textbf{SILK}, $N{=}1$ & $2^{-30}$ & \textbf{0.2} & \textbf{0.1} & \textbf{2.4} & \textbf{5.2} & \textbf{12.4} & 1 & 25.8 & 48.5 & 5.41\\
Dual MAC-256 & $2^{-26}$ & 1.76 & 1.76 & 5.1 & 10.1 & 15.8 & \textbf{0.5} & 20.2 & 42.9 & \textbf{3.62}\\
Dual MAC-512 & $2^{-26}$ & 0.88 & 0.88 & 4.8 & 9.5 & 15.2 & \textbf{0.5} & \textbf{20.1} & \textbf{42.8} & 3.63\\\midrule
\textbf{SILK}, $N{=}2$ & $2^{-61}$ & \textbf{0.2} & \textbf{0.1} & \textbf{3.2} & \textbf{5.9} & \textbf{13.4} & 1.5 & 25.8 & 48.5 & 5.41\\
MAC-256 & $2^{-58}$ & 3.1 & 3.1 & 6.1 & 11.7 & 17.7 & \textbf{0.25} & \textbf{15.9} & \textbf{38.6} & \textbf{3.67}\\MAC-512 & $2^{-58}$ & 1.6 & 1.6 & 4.4 & 8.3 & 14.0 & 0.5 & 22.9 & 44.9 & \textbf{3.67}\\\midrule
\textbf{SILK}, $N{=}4$ & $\approx2^{-122}$ & \textbf{0.2} & \textbf{0.1} & 4.6 & 7.5 & 15.5 & \textbf{2.5} & 25.8 & \textbf{48.7} & 5.41\\Seg.\ MAC$^\dagger$ & $\approx2^{-122}$ & \textbf{0.2} & \textbf{0.1} & \textbf{3.1} & \textbf{6.9} & \textbf{13.6} & 32 & \textbf{21.1} & 53.4 & \textbf{2.99}\\\bottomrule
\end{tabular}
\par
\vspace{2pt}
\noindent\parbox{\columnwidth}{\raggedright\scriptsize \emph{Notes:} The admit column reports nominal online false-acceptance bounds per key slot ($F=64$ failures), excluding offline key search and duplex corrections. A 128-bit segment tag gives approximately $2^{-122}$ per slot. Buffer size is per stream. Area and power refer to guard logic (ASAP7, 600~MHz); +SRAM includes buffers and stream table~\cite{fakeram2}. $^\dagger$Seg.\ MAC assumes ideal tag prefetch.}

\end{table}

\begin{table}[t]
\captionsetup{justification=justified,singlelinecheck=false}
\centering
\caption{RTL LLM-decoding overhead versus no guard: (a) whole-model weight reads; (b) attention with KV caches and activations.}
\label{tab:dec}
\scriptsize\setlength{\tabcolsep}{4.2pt}
\begin{tabular}{@{}l|ccc|ccc|ccc@{}}\toprule
B/cycle: 4, 8, 16$^*$ & \multicolumn{3}{c|}{Time} & \multicolumn{3}{c|}{EDP, DDR} & \multicolumn{3}{c}{EDP, HBM2}\\\midrule
\multicolumn{10}{@{}l}{\emph{(a) Whole model: Qwen2.5-0.5B weights (24$\times$7 + 19 LM-head GEMMs)}}\\
SILK & \textbf{0.41} & \textbf{0.71} & \textbf{1.07} & \textbf{2.9} & \textbf{2.6} & \textbf{2.7} & \textbf{9.2} & \textbf{8.0} & \textbf{7.4}\\
Dual MAC-256 & 1.95 & 2.24 & 2.74 & 5.6 & 5.5 & 5.8 & 10.9 & 10.3 & 10.2\\
Dual MAC-512 & 1.19 & 1.50 & 2.01 & 4.0 & 4.0 & 4.3 & \textbf{9.2} & 8.6 & 8.6\\
MAC-256 & 3.29 & 3.56 & 14.90 & 8.4 & 8.3 & 20.1 & 13.7 & 13.0 & 25.8\\MAC-512 & 1.76 & 2.06 & 7.53 & 5.2 & 5.1 & 10.6 & 10.4 & 9.8 & 15.5\\
\textit{Seg.\ MAC, commit-only} & \textit{0.25} & \textit{0.32} & \textit{0.58} & \textit{1.2} & \textit{0.9} & \textit{1.0} & \textit{3.1} & \textit{2.1} & \textit{1.6}\\
\midrule
\multicolumn{10}{@{}l}{\emph{(b) Attention: KV caches and activations (worst of 1K/4K tokens)}}\\
SILK & \textbf{0.15} & \textbf{0.47} & \textbf{1.01} & \textbf{2.4} & \textbf{2.2} & \textbf{2.4} & \textbf{9.1} & \textbf{7.9} & \textbf{7.5}\\
Dual MAC-256 & 2.65 & 3.56 & 5.14 & 6.6 & 7.2 & 8.6 & 12.4 & 12.5 & 13.7\\
MAC-256 & 3.78 & 4.67 & 17.33 & 9.0 & 9.5 & 22.8 & 14.9 & 14.9 & 29.3\\MAC-512 & 2.30 & 3.23 & 10.02 & 5.9 & 6.4 & 13.3 & 11.5 & 11.7 & 18.9\\
\textit{Seg.\ MAC, commit-only} & \textit{1.12} & \textit{2.01} & \textit{3.57} & \textit{2.3} & \textit{2.9} & \textit{4.3} & \textit{4.6} & \textit{4.4} & \textit{5.2}\\
\shortstack[l]{SILK + KV block\\table read} & \textbf{0.56} & \textbf{0.92} & \textbf{1.50} & \multicolumn{6}{c}{--}\\
\shortstack[l]{\textit{Seg.\ MAC, commit-only}\\\textit{+ KV block table read}} & \textit{7.19} & \textit{8.09} & \textit{9.68} & \multicolumn{6}{c}{--}\\

\bottomrule
\end{tabular}
\par
\vspace{2pt}
\noindent\parbox{\columnwidth}{\raggedright\scriptsize
\emph{Notes:} Values in \%; SILK holds for $N{=}1$ and $N{=}2$; strengths follow Table~\ref{tab:guards}. Dual MAC-256/512 share one runtime configuration, shown once in (b). Italics denote commit-only execution before tag verification. $^*$16~B/cycle assumes a faster Ascon engine whose timing closure has not yet been verified.}
\end{table}

Unlike weights, runtime tensors cannot be re-optimized offline, so their accuracy depends on how often check embedding must change a value. Redundant codes avoid that change; only fallback LSBs introduce an error. Runtime embedding is evaluated across eight domains (WikiText-2, C4, code, chat, math, MMLU, ARC, and HellaSwag) at $\ell=32$ and 256~B (Fig.~\ref{fig:fb}a). Fallback LSBs carry 17--37\% of INT8-activation check bits and 85.4\% of INT8-KV check bits, but only 1.9\% for NVFP4 KV, whose zero codes are more common (Fig.~\ref{fig:fb}b). INT8 KV therefore tests the calibrated fallback positions most directly. Fixed-position checks use strided or last-$\ell$ LSBs for GSM8K and last-$\ell$ LSBs for KV passkey tests (Qwen2.5 1.5B/7B; $\ell=32$). On GSM8K (900 questions per model), SILK's 95\% paired bootstrap intervals all contain zero; for INT8 KV it loses 0.4 points ($-0.3$ to $1.1$), versus 3.6 and 9.1 points for strided and last-$\ell$ positions, and fixed-position checks lose 2.1--9.1 points overall (Fig.~\ref{fig:acc}b). SILK's runtime checks change perplexity by ${\leq}0.4\%$. The benefit also holds for longer generation: SILK retrieves the passkey in all 120 trials (20 per model and length, 8K--32K tokens), whereas fixed positions miss 3 of 20 for 1.5B at 16K and 32K, and SILK changes greedy next tokens less often than INT8 quantization (Fig.~\ref{fig:acc}c). All these measurements include fallback-bit errors.

\begin{table}[t]
\captionsetup{justification=justified,singlelinecheck=false}
\centering\caption{Representative integrity and fault-protection mechanisms and their scope.}
\label{tab:related}
\scriptsize\setlength{\tabcolsep}{0.9pt}\begin{tabular}{@{}p{0.36\columnwidth}p{0.36\columnwidth}p{0.25\columnwidth}@{}}\toprule
Work & Mechanism & Scope\\
 \midrule
MGX~\cite{hua2022mgx} & Coarse-grained MACs & Accelerator memory\\SecureLoop~\cite{lee2023secureloop} & MAC-block optimization & DNN tensors\\Securator/TensorTEE~\cite{shrivastava2023securator,han2024tensortee} & Layer/tensor MACs & DNN tensors\\SeDA~\cite{xuan2025seda} & Hierarchical MACs & DNN tensors\\ProMACs~\cite{armknecht2020promacs} & Progressive MACs & Message streams\\Synergy~\cite{saileshwar2018synergy} & MACs and parity & Main memory\\RADAR~\cite{li2021radar} & Group checksums & Model weights\\HASHTAG~\cite{javaheripi2021hashtag} & Selected-layer hashes & Model weights\\In-place ECC~\cite{guan2019place,nguyen2019st} & Embedded parity & DNN memory\\E$^3$-CODE~\cite{kamineni2026e3} & Embedded ECC & Model weights\\
SANAI~\cite{rahaman2026sanai} & Embedded keyed ECC & Model weights\\Watermarks~\cite{guan2020reversible,botta2021neunac} & Embedded watermarks & Model parameters\\\midrule\textbf{SILK} & Embedded checks and chain & Weights, activations, KV\\\bottomrule
\end{tabular}
\end{table}
\subsection{Cost and Benefit of Authenticated Admission}
\label{sec:res-hw}
LLM decoding reads nearly all of its bytes in ordered streams (Fig.~\ref{fig:stream}), making external tag traffic and segment-level waits important sources of overhead. An in-order read starts at the first unit of a chain: an 8-KiB segment boundary for weights or the first token for KV. Scattered reads in the worst model consist of KV block-table lookups (4~B per page, once per layer) and token-embedding rows (one per token). Weight-segment tags add 0.20\% traffic, and runtime checks require no separate per-unit tags. The checks proceed alongside layer execution. Evicting runtime stream records incurs the protected-record traffic described in Section~\ref{sec:runtime}. Relative to the comparable Dual MAC baselines at $N=1$, SILK reduces time overhead by 47--94\% across 4–16 B/cycle streaming workloads (Tables~\ref{tab:guards} and~\ref{tab:dec}). Single-layer tests across four LLMs agree within $\pm0.25$ percentage points. At $N=2$, SILK's worst-tile time overhead is 3.2\% versus 4.4\% for MAC-512 at comparable admission strength (Table~\ref{tab:guards}). The absolute advantage grows with stream bandwidth but narrows for shorter tiles or stronger confirmation. At $N=4$, a Seg.\ MAC achieves lower worst-tile time overhead, but SILK requires only 2.5~KiB of buffering per stream versus 32~KiB for the Seg.\ MAC (Table~\ref{tab:guards}).

At 4--8~B/cycle, lower traffic cuts energy-delay product (EDP) overhead versus the Dual MAC baselines by 28--69\% with DDR and up to 37\% with HBM2. HBM2 offers less benefit because memory transfers are cheaper, so SILK's additional logic power matters more. SILK uses about 28\% more guard-logic area than Dual MAC-256. Buffering instead favors SILK against segment-level admission: at $N=4$, SILK occupies 48.7k~\textmu m$^2$ including SRAM, versus 53.4k~\textmu m$^2$ for the Seg.\ MAC (Table~\ref{tab:guards}). At $N=1$, the total is 48.5k~\textmu m$^2$, or 25\% of the accelerator.

Finally, a commit-only Seg.\ MAC illustrates what is gained by enforcing verification before use. It computes before checking its tag and can therefore be faster on uninterrupted weight streams, but it does not provide the same admission guarantee. SILK adds at most 0.5 percentage points of time overhead beyond this reference on weights. Dependent reads reverse this ranking: a KV block-table lookup (scattered accesses in Fig.~\ref{fig:stream}) names the next request, which cannot issue until the table is admitted. We model each layer's block table as a protected 512-B object whose entries leave the cover-bit positions free for checks; the next request waits for the table's confirm event in SILK and for the tag of its 8-KiB segment in the commit-only Seg.\ MAC. At 4~B/cycle, these lookups add 0.41 percentage points of time overhead for SILK and 6.07 for the Seg.\ MAC, a 14.8$\times$ ratio that remains 12--15$\times$ across 4--16~B/cycle (Table~\ref{tab:dec}b). For the per-unit MACs of Table~\ref{tab:dec}b, block-table reads add 0.29--0.52 percentage points to their overheads, which then reach 2.6--17.8\% at 4--16~B/cycle, versus 0.56--1.50\% for SILK. The benefit depends on stream access: a random 512-B weight object transfers 8.7$\times$ its bytes and takes 5.3$\times$ MAC-256's time.

\section{Conclusion}
\label{sec:conclusion}\silk{} combines quantization-aware check embedding, delayed-absorption authentication, and guarded commit for streaming LLM inference. Across the evaluated symmetric INT4 models, mean perplexity increases remain below 0.4\% with 0.2\% weight metadata. The RTL results show lower authentication overhead on ordered streams, with tradeoffs in guard area, buffering, and confirmation depth. These tradeoffs define the operating conditions in which in-place checks benefit an accelerator.

\clearpage
\bibliographystyle{IEEEtran}
\bibliography{references}

\end{document}